# Transmitter Diversity with Beam Steering

Osama Zwaid Alsulami[1], Mohammed T. Alresheedi[2] and Jaafar M. H. Elmirghani[1]
[1]School of Electronic and Electrical Engineering, University of Leeds, LS2 9JT, United Kingdom
[2]Department of Electrical Engineering, King Saud University, Riyadh, Kingdom of Saudi Arabia
**ml15ozma@leeds.ac.uk, malresheedi@ksu.edu.sa, j.m.h.elmirghani@leeds.ac.uk**

**ABSTRACT**
Providing high data rates is one of the drivers in visible light communication (VLC) systems. This paper introduces a VLC system that employs angle diversity transmitters with beam steering to provide high data rates. In this work, red, yellow, green, and blue (RYGB) laser diodes (LD) are used as transmitters to obtain a high modulation bandwidth. Two types of RYGB LDs units are used in this paper: angle diversity transmitters (ADT) RYGB LDs light unit (for illumination and communication) and RYGB LDs light unit (for illumination). In addition, a four branch angle diversity receiver (ADR) is used where we study the delay spread and SNR. The proposed system was compared to the normal VLC system. A data rate up to 22.8 Gb/s was achieved using simple on-off-keying (OOK) modulation.

## 1. INTRODUCTION

Due to the growth in demand for high data rates, visible light communication (VLC) has received interest for transferring data. Currently, radio frequency (RF) is used to carry data especially indoors. However, traditional radio communication systems suffer from limited channel capacity and transmission rate due to the limited radio spectrum available, while the data rates requested by the users continue to increase exponentially. Achieving very high data rates (beyond 10 Gbps and into the Tbps regime) using the bandwidth of radio systems is challenging. According to Cisco, mobile Internet traffic over this half of the decade (2016-2021) is expected to increase by 27 times [1]. Given this expectation of dramatically growing demand for data rates, the quest is already underway for alternative spectrum bands beyond radio waves. The latter are bands currently used and planned for near future systems, such as 5G cellular systems [2]. Optical wireless (OW) systems and VLC systems can provide license free bandwidth, high security and low-cost compared to RF systems [3]-[10]. However, VLC and OW systems have some limitations and one of these limitations is the potential absence of line-of-sight (LOS) components in the link which impairs the system's performance significantly [11], [12]. In addition, inter-symbol interference (ISI) caused by multi-path propagation in VLC systems can affect the system's performance. VLC systems have been shown in many studies to be able to transmit video, data and voice, at data rates up to 20 Gbps in indoor systems [13]-[18].

In this paper, a VLC system based on angle diversity transmitters with beam steering is introduced in conjunction with RYGB LDs and an angle diversity receiver (ADR). The RYGB LDs can provide a white colour which can be used for indoor illumination as in [19]. The RYGB LDs were used as transmitters to provide a high modulation bandwidth and a ADR is utilised as an optical receiver to reduce the effect of ISI which can enhance the performance of the system. The effects of mobility and multi-path propagation are considered in this paper. This paper is organised as follows: The room configuration is described in Section 2. The optical receiver design is given in Section 3. Section 4 provides the optical transmitter design. Section 5 shows the simulation results and the conclusions are drawn in Section 6.

## 2. ROOM CONFIGURATION

As shown in Figure 1 the room used in the simulations has dimensions (length × width × height) of 8 m × 4 m × 3 m. An empty room has been used in the simulation that has no doors or windows. Reflections up to second order were considered in this work since the third and higher order reflections have no impact on the received optical power [20], [21]. A ray-tracing algorithm was utilised for modelling reflections from the ceiling, walls and the floor in the room. Thus, each surface in the room was divided into small equal areas of size $dA$ with a reflection coefficient of $\rho$. The authors in [21] showed that plaster walls reflect light rays close to a Lambertian pattern. Therefore, each surface in the room such as ceiling, walls and floor was modelled as a Lambertian reflector with reflection coefficient equal to 0.8 for ceiling and walls and 0.3 for the floor. Each element in each surface acts as a small emitter that reflects the received ray in the form of a Lambertian pattern with $n$ (emission order of the Lambertain pattren) equal to 1. The area of the surface elements determines the resolution of the results. When the surface elements are very small, higher time resolution results are achieved, for example when evaluating the impulse response. However, this improved resolution comes at the expense of long computation time in the simulation. Therefore, 5 cm × 5 cm was chosen as an area of the surface element for the first order reflection, while 20 cm × 20 cm was chosen as an area of the surface element for the second order reflection to keep the computation time of the simulation within a reasonable time [15], [22]. The communication floor (CF) was set as 1 m above the floor as shown in Figure 1 which means all communications are done above the CF.

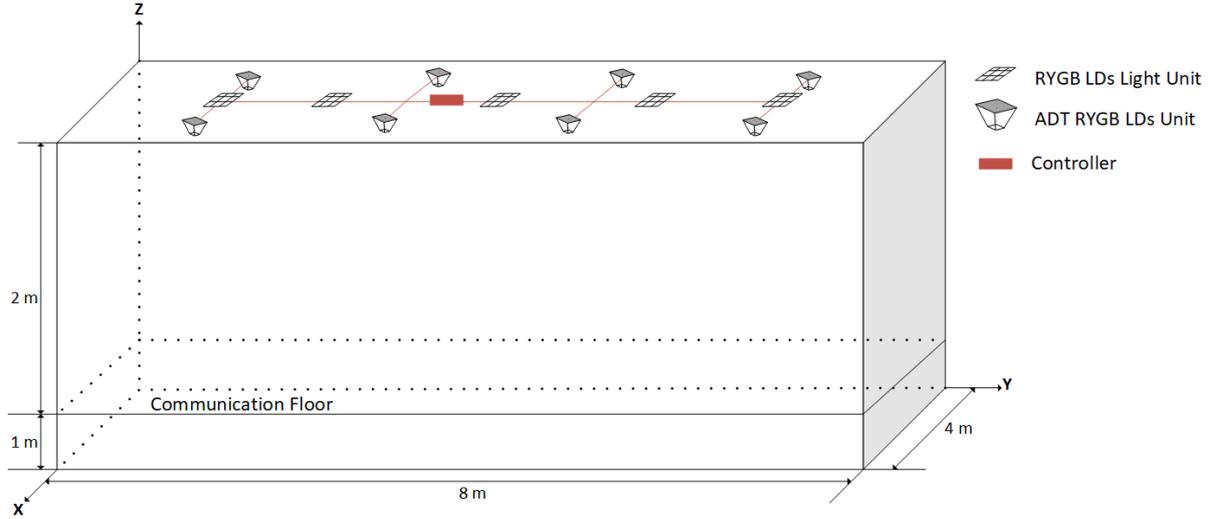

Figure 1: Room Configuration [23]-[25]

## 3. OPTICAL RECEIVER DESIGN

In this paper, a four branch ADR (see Figure 2) with narrow FOVs was used to collect signals and reduce the inter-symbol interference (ISI). The design is based on the ADRs in [12], [13], [23], [26]. Each photodetector in the ADR is oriented to different direction to cover different areas in the room based on two angles: Azimuth ($Az$) and Elevation ($El$). The $El$ angles of the detectors are set equal to 70°. While, the $Az$ angles of these detectors have been chosen to be 45°, 135°, 225° and 315°. In addition, a narrow FOV has been utilised that is equal to 21° for all detectors to reduce the ISI. The area of each photodetector was chosen equal to $4\ mm^2$ with responsivity equal to 0.4 A/W. The receiver was examined along $x$ = 2m and different $y$-axis locations, namely $y$ = 1 m, 2 m, 3 m, 4 m, 5 m, 6 m and 7 m.

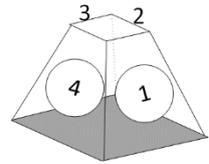

Figure 2: Optical Receiver Design

## 4. OPTICAL TRANSMITTER DESIGN

In this paper, two types of light units were used. The first is a light unit that provides VLC and supports angle diversity transmitter (ADT) using RYGB and provides illumination also. The second type of light units uses RYGB LDs just for illumination.

### 4.1 ADT RYGB LDs light unit Design

This paper uses the angle diversity transmitter concepts which were studied in [23]-[25] and introduces beam steering with these ADTs for the first time. Each ADT RYGB LDs light unit consist of 4 branches (see Figure 3). Each one of these branches consist of three narrow beam RYGB LDs that cover a small different area inside the room. The size of the beam used depends on the beam Lambertian emission order ($n$). Thus, when $n$ is increased the beam size is reduced and the coverage area of the RYGB LD is also reduced which can reduce the intersection and interference between adjacent branches. However, when the intersection between adjacent branches becomes zero, gaps will appear between the areas covered by these branches and that results in reducing the average SNR achieved over the entire room and hence the performance of the communication system is impaired. Therefore, the half beam angle of each RYGB LD is set at 21º to reduce gaps between adjacent branches. The light units are located on the ceiling at different locations, namely at: (1 m, 1 m, 3 m), (1 m, 3 m, 3 m), (1 m, 5 m, 3 m), (1 m, 7 m, 3 m), (3 m, 1 m, 3 m), (3 m, 3 m, 3 m), (3 m, 5 m, 3 m) and (3 m, 7 m, 3 m) as shown in Figure 1. Each branch of each light unit is oriented to different area by using two angles $El$ and $Az$. The $El$ angles of these branches were set equal to -70⁰. The *Az* angles of these branches have been chosen to be 45°, 135°, 225° and 315°. In addition, an optical hologram similar to [27] was used in front of each branch to provide beam steering with the ADR [28], [29], (with possible extension to imaging receivers [30]). As a result, the received power increased.

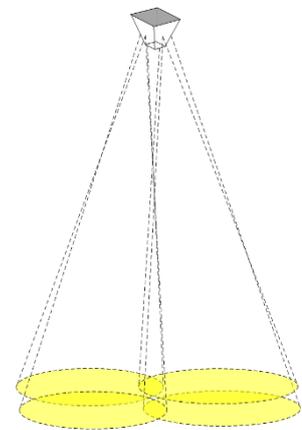

Figure 3: Optical Transmitter Design

**4.2 Hologram design and beam steering in ADT unit**

Using a hologram can increase the received power and reduce the ISI. Thus, by using a hologram with beam steering, the number of transmitter faces can be reduced and a similar communication performance can be achieved. The basic idea of the hologram is as follows:

- A signal is sent by each ADT branch individually starting by the first branch in the first unit to select the best branch based on the SNR.
- Then, the coverage area of the selected branch is divided into four quadrants depend on its transmission angles, which are from -21⁰ to 21, to choose the best quadrant that has a higher SNR as a new coverage area.
- Subsequently, the new coverage area is divided into four sub-quadrants and this process is repeated until the receiver location is identified.
- Then, the beam is steered to the receiver location.

**4.3 RYGB LDs light unit Design**

Acceptable lighting levels inside the room based on the ISO and European illumination requirements cannot be achieved when just using the ADT light units [31]. Thus, five RYGB Light units were utilised to provide acceptable lighting level inside the room. Each unit of these RYGB Light units has 9 wide-semi angles RYGB LDs. The semi angle of the RYGB LDs was chosen to be 70⁰ to increase the lighting level inside the room. RYGB LDs light units are located on the ceiling in different locations: (2 m, 1 m, 3 m), (2 m, 2.5 m, 3 m), (2 m, 4 m, 3 m), (2 m, 5.5 m, 3 m) and (2 m, 7 m, 3 m) as shown in Figure 1. Figure 4 illustrates the illumination (lx) inside the room using one branch from ADT RYGB LDs light units as a beam steering source. A minimum illumination level equal to 313.7 lx was achieved which is above the minimum requirement for illumination in the room based on the ISO and European illumination requirements. The maximum allowed illumination level was not exceeded.

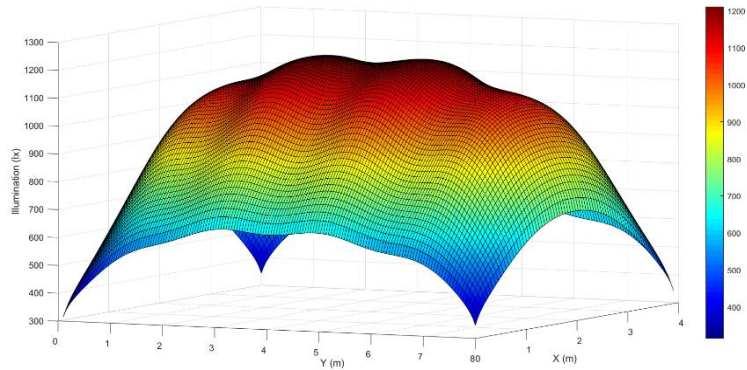

Figure 4: The illumination (lx) inside the room

**5. SIMULATION SETUP AND RESULTS**

In this work, the simulation starts by choosing the best branch of ADT RYGB LDs light units based on SNR. For identifying the best branch for transferring data, each branch of each ADT RYGB LDs light unit has a unique ID. Thus, an algorithm is used to identify the best branch by sending a signal through the first branch in the first ADT RYGB LDs light unit. At the receiver, the SNR is calculated and transferred to the controller through a low infrared data rate system. This step is repeated with each branch to choose the best branch. The specific location of the receiver is then identified by repeating the process of dividing the coverage area of the selected branch into four quadrants until the coverage area become 10 cm × 10 cm. Subsequently, 50% of the whole power of the chosen branch is oriented to the receiver location using beam steering to provide high receiver SNR. The proposed system was compared with a normal VLC system that uses 8 RYGB LDs light units for illumination and communication placed on the ceiling at different locations: (1 m, 1 m, 3 m), (1 m, 3 m, 3 m), (1 m, 5 m, 3 m), (1 m, 7 m, 3 m), (3 m, 1 m, 3 m), (3 m, 3 m, 3 m), (3 m, 5 m, 3 m) and (3 m, 7 m, 3 m) using the same ADR design. The proposed system reduced the delay spread significantly from 0.0121 ns to 0.000122 ns when the receiver is placed at the room centre (2 m, 4 m, 1 m) as shown in Figure 5a. The SNR was calculated based on the best ADR branch. The proposed system provides a significant improvement in SNR compared to the normal VLC system, enabling performance as shown in Figure 5b at high data rate up to 22.8 Gb/s for OOK.

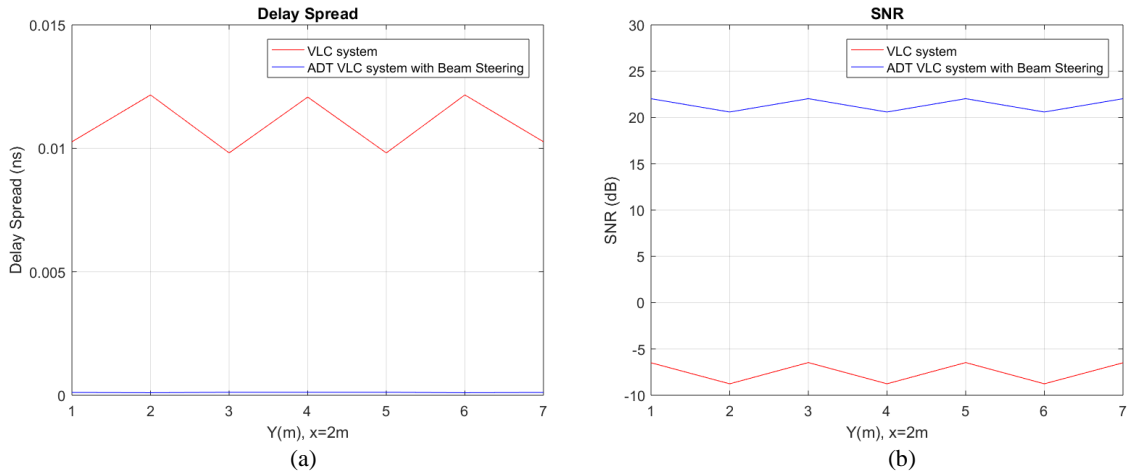

Figure 5: (a) Delay Spread, (b) SNR

## 6. CONCLUSIONS

A VLC system was proposed in this paper that uses angle diversity transmitters. It can offer a high data rate up to 22.8 Gb/s over the entire room while using a simple modulation format, (OOK). This VLC system used two types of RYGB LDs units: ADT RYGB LDs light units (for illumination and communication) and RYGB LDs light units (for illumination). In addition, a four branch ADR was used and the delay spread and SNR were determined. The proposed system was compared to a normal VLC system. It provides a significant improvement in the delay spread and SNR and supports operation at high data rate of up to 22.8 Gb/s.

## ACKNOWLEDGEMENTS

The authors would like to acknowledge funding from the Engineering and Physical Sciences Research Council (EPSRC) for the TOWS project (EP/S016570/1). All data are provided in full in the results section of this paper.